\documentclass[epj]{webofc}
\usepackage[utf8]{inputenc}
\usepackage[varg]{txfonts}   
\usepackage{booktabs}
\usepackage{xcolor}
\definecolor{darkred}{rgb}{0.4,0.0,0.0}
\definecolor{darkgreen}{rgb}{0.0,0.4,0.0}
\definecolor{darkblue}{rgb}{0.0,0.0,0.4}
\usepackage[bookmarks,linktocpage,colorlinks,
    linkcolor = darkred,
    urlcolor  = darkblue,
    citecolor = darkgreen]{hyperref}
\usepackage{subfigure}
\wocname{EPJ Web of Conferences}
\woctitle{Lattice2017}
\newcommand{\e}{{\rm{e}}}
\newcommand{\img}{{\rm{i}}}

\def\slashchar#1{\setbox0=\hbox{$#1$} 
\dimen0=\wd0 
\setbox1=\hbox{/} \dimen1=\wd1 
\ifdim\dimen0>\dimen1 
\rlap{\hbox to \dimen0{\hfil/\hfil}} 
#1 
\else 
\rlap{\hbox to \dimen1{\hfil$#1$\hfil}} 
/ 
\fi}

\begin{document}
%
\selectlanguage{english}
\title{
Dispersion relation and unphysical poles of M\"obius domain-wall fermions in free field theory at finite $L_s$
}
\author{%
\firstname{Masaaki} \lastname{Tomii}\inst{1}\fnsep\thanks{Speaker, \email{mt3164@columbia.edu}, supported in part by US DOE grant \#DE-SC0011941}
}
\institute{%
Physics Department, Columbia University, New York 10027, USA
}
\abstract{%
We investigate the dispersion relation of M\"obius domain-wall fermions in free field theory at finite $L_s$. We find that there are $L_s-1$ extra poles of M\"obius domain-wall fermions in addition to the pole which realizes the physical mode in the continuum limit. The unphysical contribution of these extra poles could be significant when we introduce heavy quarks. We show in this report the fundamental properties of these unphysical poles and discuss the optimal choice of M\"obius parameters to minimize their contribution to four-dimensional physics.
}
\maketitle
\section{Introduction}\label{intro}
\label{sec:intro}

Lattice calculation including the charm quark as well as the lighter quarks is
desired to give accurate prediction of the Standard Model, which could play
a key role in probing for new physics beyond the Standard Model.
Since the charm-quark mass is comparable to currently available lattice cutoffs,
the charm quark on the lattice could induce significant discretization errors.

Domain-wall fermions with large input masses are known to have some special
difficulties as well as the na\"\i ve $O(a)$ discretization errors.
Such difficulties were originally suggested \cite{Liu:2003kp,Christ:2004gc}
by analyzing the eigenvalues of the hermitian version of the domain-wall
operator, the five-dimensional Dirac operator multiplied by the chirality operator
$\gamma_5$ and the five-dimensional reflection operator.
That work explained that the hermitian operator involves unphysical modes as well
as the physical modes and that the eigenvalues of unphysical modes are largely
independent of the input quark mass, while those of physical modes are roughly
proportional to the input mass.
This implies that the dominance of physical modes would be lost as
the input mass increases.

A few years later, \cite{Dudek:2006ej} observed the oscillatory behavior of
correlation functions, which is known as a particular issue of domain-wall fermions
and is observed at large values of the domain-wall height such as
$M_5=1.7$.
This oscillatory behavior was described as the result of negative eigenvalues
of the transfer matrix \cite{Syritsyn:2007mp}, which were found to exist at large
values of $M_5$, $M_5>1$ in the case of free field theory.

In this work, we propose another point of view to describe such a curious
artifact through an investigation of the pole structure of M\"obius domain-wall
fermions in free field theory.
Although a study of the pole structure of the domain-wall fermion propagator was recently
done \cite{Liang:2013eoa,Sufian:2016cft}, we find this previous work contains some
mistakes in the analytic formula of the quark propagator and the understanding of the
pole structure.
We provide corrections to these mistakes and discuss the dependence of the pole
structure on the domain-wall height $M_5$ and the difference between the M\"obius
parameters $b-c$.

This article is organized as follows.
In Section~\ref{sec:qprop}, we give definitions and the propagators of M\"obius
domain-wall fermions both in four and five dimensions.
In Section~\ref{sec:unphys_poles}, the presence of unphysical poles of domain-wall
fermions is demonstrated.
In Section~\ref{sec:disp_rel}, we show the energy-momentum dispersion relations for
the physical and unphysical poles and discuss their dependence on the parameters
of M\"obius domain-wall fermions.

\section{Definitions and propagator of M\"obius domain-wall fermions at finite $L_s$}
\label{sec:qprop}

We consider the M\"obius domain-wall fermion action $\overline \psi D_{\rm MDW}\psi$,
where the Dirac operator $D_{\rm MDW}$ in momentum space is given by
\begin{equation}
(D_{\rm MDW})_{s,t}
= \tilde D\delta_{s,t}
-(P_+\delta_{s,t+1}+P_-\delta_{s,t-1})
+m(P_+\delta_{s,0}\delta_{t,L_s-1} + P_-\delta_{s,L_s-1}\delta_{t,0}).
\end{equation}
Here, we use the chiral projection operators
$P_\pm = \frac{1}{2}(1\pm\gamma_5)$ and
\begin{equation}
\tilde D = D_-^{-1}D_+,\ \ 
D_+ = 1+bD_W,\ \ 
D_- = 1-cD_W,
\end{equation}
with the Wilson Dirac operator $D_W$ at a negative mass parameter $-M_5$,
\begin{equation}
D_W = \img\slashchar{\tilde p}+\sum_\mu(1-\cos p_\mu) -M_5,
\end{equation}
where $\slashchar{\tilde p} = \sum_\mu\gamma_\mu\sin p_\mu$.
For simplicity, we omit the lattice spacing $a$ and express everything
in lattice units throughout this article.
As is shown in \cite{Brower:2012vk,Blum:2014tka},
if the four-dimensional quark fields are defined as
\begin{equation}
q = P_-\psi_0+P_+\psi_{L_s-1},\ \ 
\overline q = \overline\psi_0P_+ + \overline\psi_{L_s-1}P_-,
\end{equation}
the corresponding quark propagator has the form
\begin{align}
S_F^{4d}(p)
&= P_-(D_{\rm MDW}^{-1})_{0,0}P_+
+P_+(D_{\rm MDW}^{-1})_{L_s-1,L_s-1}P_-
\notag\\& \hspace{4mm}
+P_-(D_{\rm MDW}^{-1})_{0,L_s-1}P_-
+ P_+(D_{\rm MDW}^{-1})_{L_s-1,0}P_+,
\label{eq:4Dquarks}
\end{align}
which is the same as the propagator of overlap fermions up to an overall factor
and a contact term in the limit of infinite $L_s$.
Besides the mass parameter, there are four input parameters:
the extent of the fifth dimension $L_s$, the domain-wall height $M_5$, the
M\"obius parameters $b$ and $c$.
Although these parameters characterize the regularization and do not affect
any observables in the continuum limit after fermion mass renormalization,
discretization errors at finite lattice spacings depend on them.

We can rewrite the five-dimensional Dirac operator $D_{\rm MDW}$ as
\begin{equation}
D_{\rm MDW} = {b+c\over D_-^\dag D_-}\img\slashchar{\tilde p} + W^+P_- + W^-P_+,
\end{equation}
where we define
\begin{align}
W^\pm_{s,t} &= W\delta_{s,t}-\delta_{s\pm1,t}+m\delta_{s/t,L_s-1}\delta_{t/s,0},
\label{eq:Wpm_st}
\\
W &= {-bc(\tilde p^2+{\cal M}^2)+(b-c){\cal M}+1\over D_-^\dag D_-},
\\
{\cal M}&=\sum_\mu(1-\cos p_\mu) -M_5,
\\
D_-^\dag D_- &= c^2(\tilde p^2+{\cal M}^2)-2c{\cal M}+1,
\end{align}
with $\tilde p^2 = \sum_\mu\sin^2p_\mu$.
The five-dimensional propagator of M\"obius domain-wall fermions
can thus be obtained in the same manner \cite{Shamir:1993zy}
as for Shamir domain-wall fermions:
\begin{align}
D^{-1}_{\rm MDW}
&= \left[-{b+c\over D_-^\dag D_-}\img\slashchar{\tilde p}+W^-\right]G^-P_-
+ \left[-{b+c\over D_-^\dag D_-}\img\slashchar{\tilde p}+W^+\right]G^+P_+,
\\
G^\pm_{s,t} &= \left[\bigg({b+c\over D_-^\dag D_-}\bigg)^2\tilde p^2+W^\mp W^\pm\right]^{-1}_{s,t}
\notag\\
&= A_0\e^{-\alpha|s-t|}+A_\pm\e^{\alpha(s+t-L_s+1)}+A_\mp\e^{-\alpha(s+t-L_s+1)}
+A_m\cosh[\alpha(s-t)],
\label{eq:G_comp}
\\
\cosh\alpha &=
\frac{\big({b+c\over D_-^\dag D_-}\big)^2\tilde p^2 + W^2 + 1}{2W},
\\
A_0 &= \frac{1}{2W\sinh\alpha},
\\
A_\pm &= \frac{A_0}{F_{L_s}}(1-m^2)(W-\e^{\mp\alpha}),
\label{eq:Apm}
\\
A_m &= \frac{A_0}{F_{L_s}}
\left[
4mW\sinh\alpha-2(W\e^{-\alpha}-1+m^2(1-W\e^\alpha))\e^{-\alpha L_s}
\right],
\label{eq:Am}
\\
F_{L_s}
&= \e^{\alpha L_s}(1-W\e^\alpha+m^2(W\e^{-\alpha}-1))-4mW\sinh\alpha
\notag\\&\hspace{4mm}
+\e^{-\alpha L_s}(W\e^{-\alpha}-1+m^2(1-W\e^\alpha)).
\label{eq:FLs}
\end{align}

Inserting this result into \eqref{eq:4Dquarks}, we obtain
\begin{align}
S_F^{4d}(p)
&= {2\sinh(\alpha L_s)\over F_{L_s}}{b+c\over D_-^\dag D_-}
\img\slashchar{\tilde p}
\notag\\&\hspace{4mm}
+{2\over F_{L_s}}
\left\{
m[W\sinh(\alpha(L_s-1))-\sinh(\alpha L_s)]
-W\sinh\alpha
\right\}.
\label{eq:qprop_4d}
\end{align}

\section{Unphysical poles at finite $L_s$}
\label{sec:unphys_poles}

As is well known, the continuum limit of a lattice fermion reproduces
only a relevant Dirac field as long as doublers have been removed.
Actual lattice calculations are however carried out at finite lattice spacings,
where unphysical extra modes could appear depending on the details of
lattice action.
Such poles may induce complicated discretization errors which are not
easily controlled.
In this section, we show the presence of unphysical poles of M\"obius
domain-wall fermions in free field theory.

\begin{figure}[tbp]
\begin{center}
\subfigure{\mbox{\raisebox{1mm}{\includegraphics[width=86mm, bb=0 0 345 230]{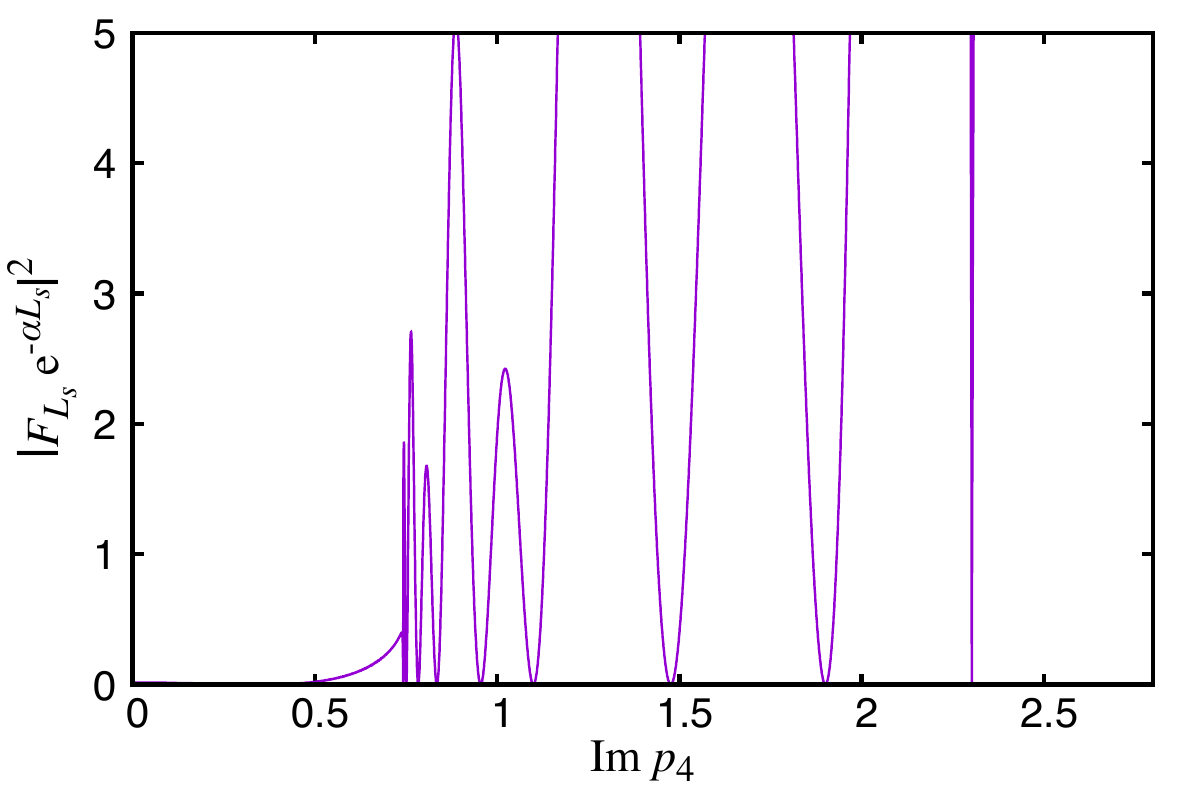}}}}
\subfigure{\mbox{\raisebox{1mm}{\includegraphics[width=86mm, bb=0 0 345 230]{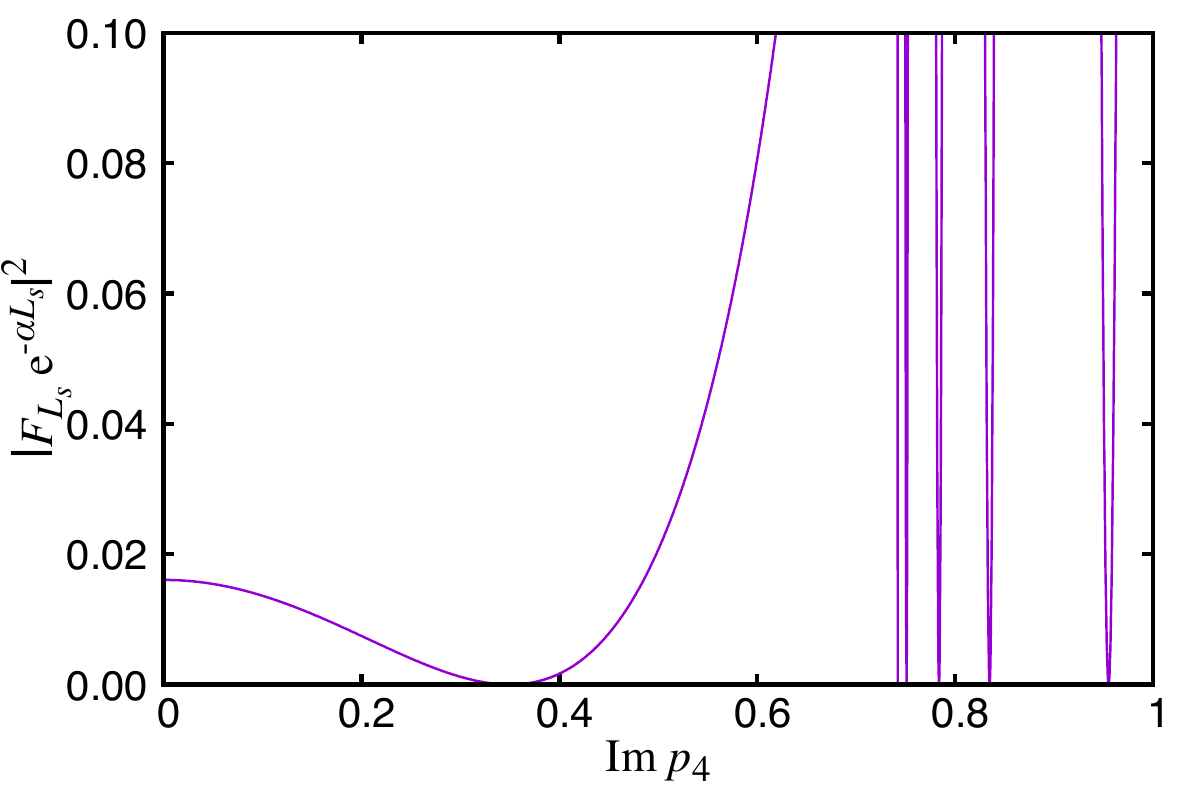}}}}
\subfigure{\mbox{\raisebox{1mm}{\includegraphics[width=86mm, bb=0 0 345 230]{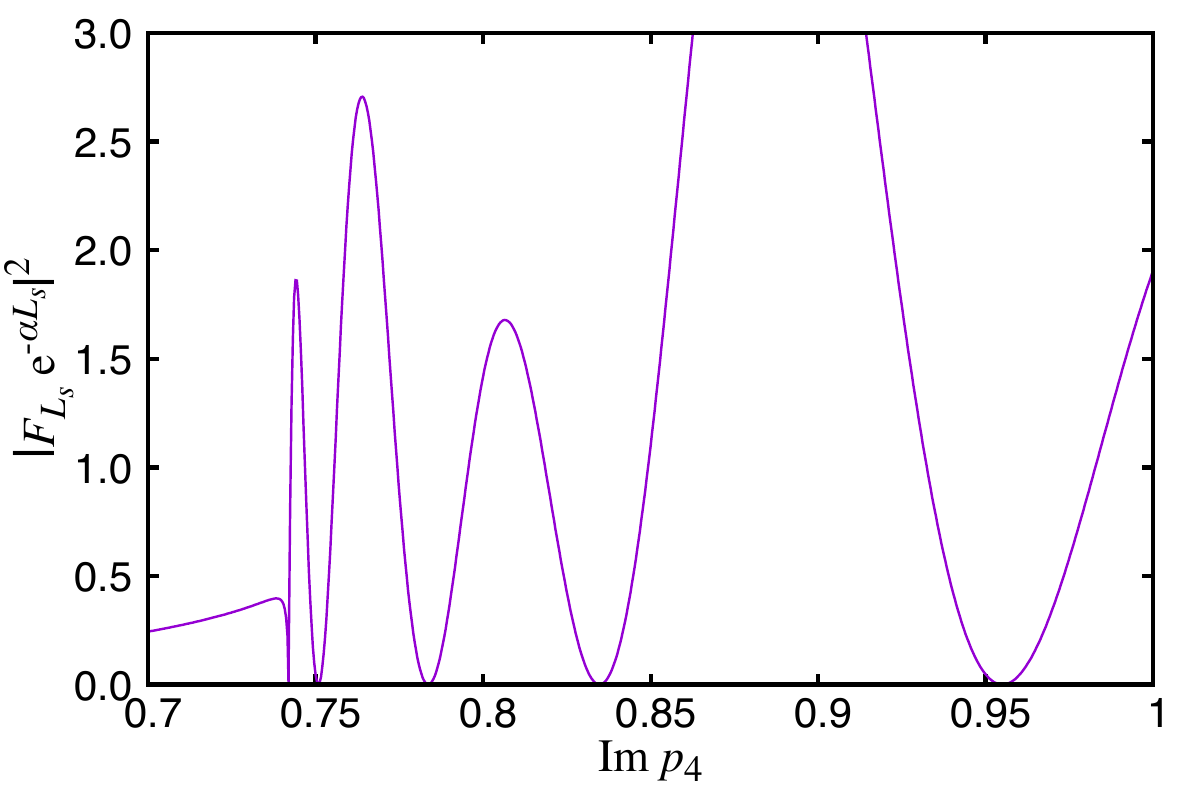}}}}
\caption{
$|F_{L_s}\e^{-\alpha L_s}|^2$ calculated at $L_s=8$, $M_5=0.9$, $m_f^{\rm pole}=0.35$,
$b-c=1$, $b+c=1$, $\vec p = 0$ and ${\rm Re}~p_4 = 0$ plotted as a function of
${\rm Im}~p_4$.
The lower panels show some magnifications of complicated parts in the top panel,
which accommodates all the zero points of $F_{L_s}$.
}
\label{fig:FLs}
\end{center}
\end{figure}

Figure~\ref{fig:FLs} shows $|F_{L_s}\e^{-\alpha L_s}|^2$ calculated at
$b=1$, $c=0$, $L_s=8$, $M_5=0.9$,
$\vec p=0$ and ${\rm Re}~p_4=0$.
Here, the input mass parameter $m$ is tuned so that the physical pole
mass $m_f^{\rm pole}$ reads 0.35.
The lower panels show some magnifications of complicated parts in the
upper panel, which accommodates all the zero points of $F_{L_s}$.
Since we find all of the zero points are located on the imaginary axis of
$p_4$ for this parameter choice, we plot the result only at ${\rm Re}~p_4=0$.
While the physical pole is seen at ${\rm Im}~p_4=m_f^{\rm pole}=0.35$,
there are nine other zero points of $F_{L_s}$.
Two of them are identified as the solutions of $\cosh\alpha=1$ or
$\cosh\alpha=-1$, which correspond to ${\rm Im}~p_4 \simeq 2.30$
and 0.74 in the plot, respectively.
The other seven zero points are located between these two special zero points.

The special two zero points satisfying $\cosh\alpha = \pm1$ are not poles of
the propagator \eqref{eq:qprop_4d} since the numerator of the propagator also
vanishes at these points and the limit $\lim_{\alpha\rightarrow0,\img\pi} S_F^{4d}(p)$
is still finite.
On the other hand, the quark propagator at each of the remaining seven zero
points is singular, indicating these zero points are unphysical poles.
These unphysical poles are located in the region $-1<\cosh\alpha<1$,
where $\alpha$ is pure imaginary and therefore any terms in \eqref{eq:FLs}
are not suppressed at large values of $L_s$, showing some oscillations
with varying ${\rm Im}~p_4$.
Since the number of these oscillations is proportional to $L_s$,
the number of unphysical poles increases as $L_s$ increases and
we find $L_s-1$ unphysical poles in our analysis.

\section{Energy-momentum dispersion relation}
\label{sec:disp_rel}

As is well known, a pole at $p_4 = iE$ in Euclidean space
behaves $\sim \e^{-Et}$ in coordinate space.
If the energy $E$ of the physical pole is close to or larger than that of unphysical
poles, the signal of the physical pole may be unclear even at long distances.

The relation between the spatial momentum and the pole energy is represented
by the energy-momentum dispersion relation.
The dispersion relation on the lattice deviates from that in the continuum
limit with $O(a^2)$ error allowing $O(4)$-violating terms.
The dispersion relations for improved overlap fermions using the Brillouin kernel
were investigated \cite{Cho:2015ffa,Durr:2017wfi}.
In this section, we concentrate on unimproved M\"obius domain-wall fermions
and show the dispersion relation for both the physical and unphysical poles.

\begin{figure}[t]
\begin{center}
\includegraphics[width=90mm, bb=0 0 345 230]{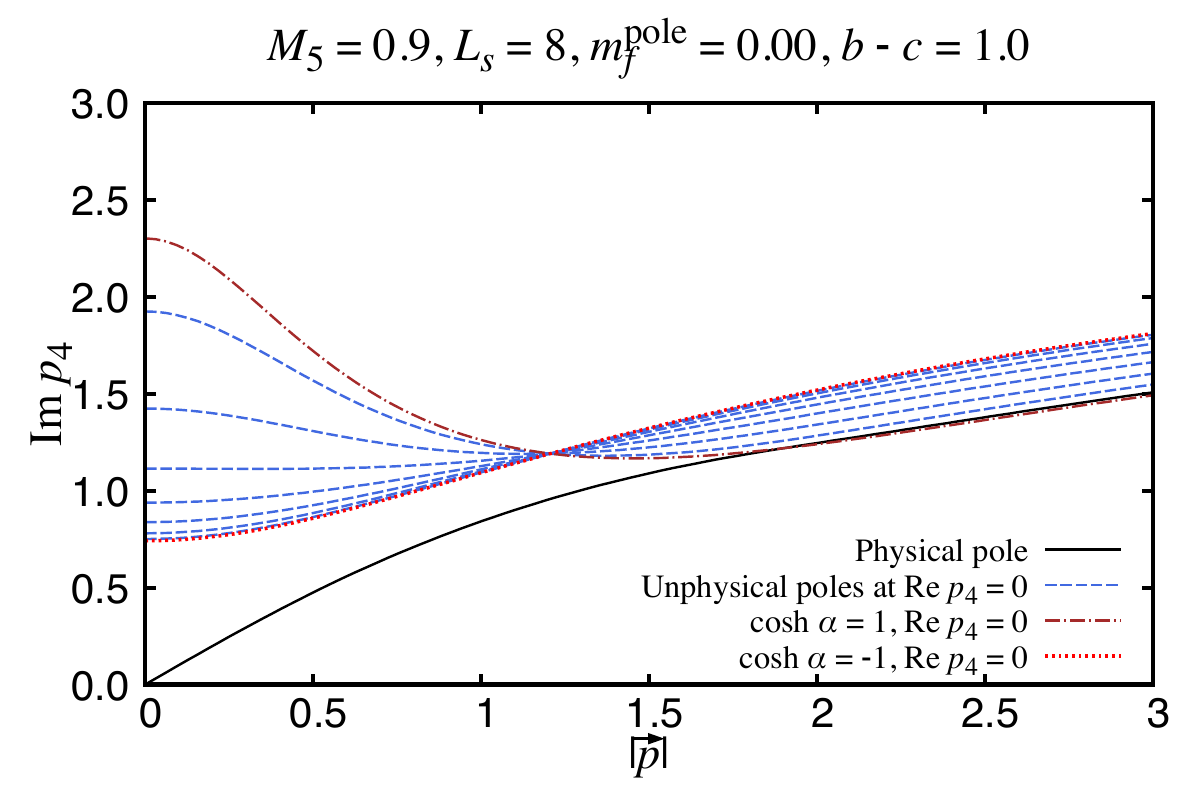}
\caption{
Dispersion relation for the domain-wall fermion at
$M_5=0.9, L_s=8, m_f^{\rm pole}=0, b+c=1, b-c=1$ and spatial momentum
$\vec p = ({|\vec p\,|\over\sqrt{3}}, {|\vec p\,|\over\sqrt{3}}, {|\vec p\,|\over\sqrt{3}})$.
}
\label{fig:UPP_M50.9}
\end{center}
\end{figure}

Figure~\ref{fig:UPP_M50.9} shows the dispersion relation for domain-wall
fermions at $b=1, c=0$, $M_5 = 0.9$, $L_s=8$ and $m_f^{\rm pole}=0$.
We choose the spatial momentum in the diagonal direction,
$\vec p = ({|\vec p\,|\over\sqrt{3}}, {|\vec p\,|\over\sqrt{3}}, {|\vec p\,|\over\sqrt{3}})$.
At any spatial momenta, there are one physical (solid curve) and seven unphysical
poles (dashed curves) on the imaginary axis of $p_4$.

As discussed in the previous section, these unphysical poles are located in
the region between two curves, $\cosh\alpha=1$ (dashed-dotted curve) and
$\cosh\alpha=-1$ (dotted curve).
The boundaries $\cosh\alpha=\pm1$ are analytically given by
\begin{align}
\cos p_4|_{\cosh\alpha=1} &= \frac{\sum_{i=1}^3\sin^2p_i+B^2+1}{2B},
\label{eq:sol_cosha_p1}
\\*
\cos p_4|_{\cosh\alpha=-1}
&= \frac{4+4(b-c)B+(b-c)^2(\sum_{i=1}^3\sin^2p_i+B^2+1)}{4(b-c)+2(b-c)^2B},
\label{eq:sol_cosha_m1}
\\
B &= 4 -M_5 - \sum_{i=1}^3\cos p_i.
\end{align}
Note that the solution of $\cosh\alpha=1$, the lower or upper bound
on the unphysical poles, depends only on $M_5$ and $p_i$, while the other
bound, the solution of $\cosh\alpha=-1$, depends also on $b-c$.
This fact motivates us to vary $b-c$ as well as $M_5$, although $b-c$
has not usually been tuned to minimize discretization errors.

\begin{figure}[tbp]
\begin{minipage}{0.51\hsize}
\includegraphics[width=70mm, bb=0 0 345 230]{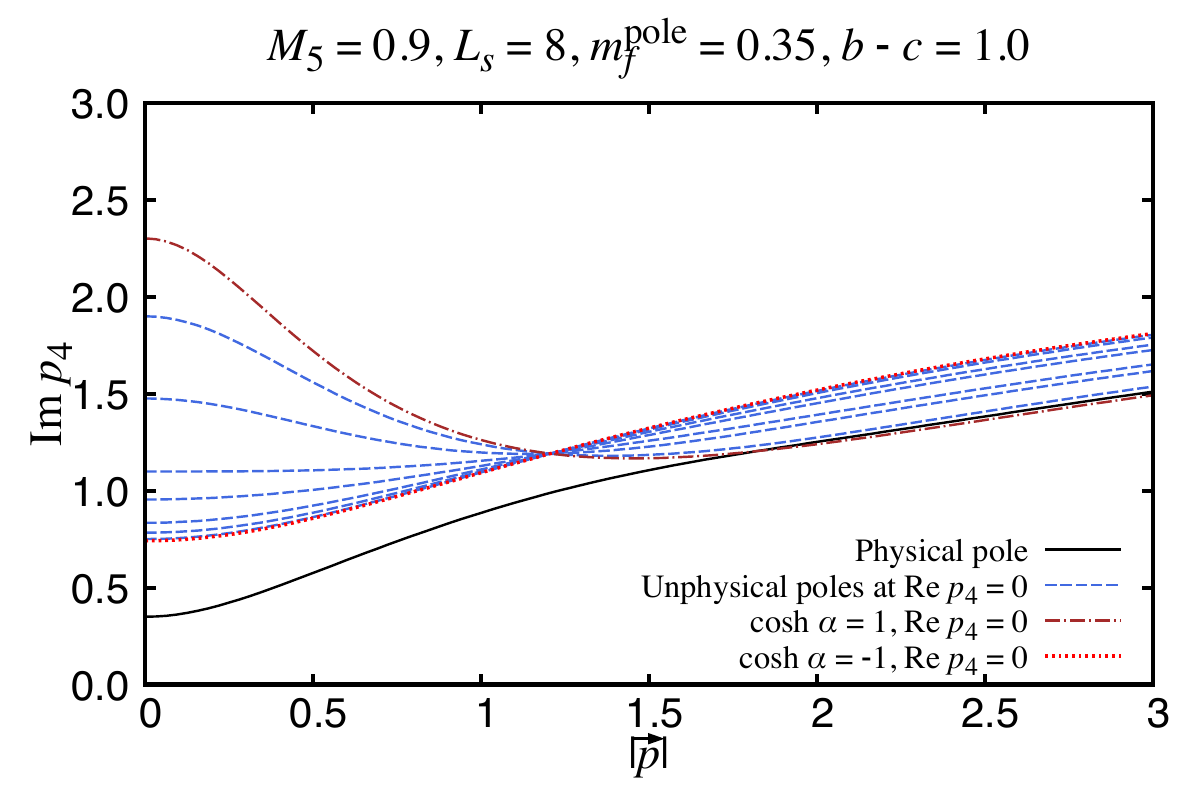}
\caption{
Same as Figure~\ref{fig:UPP_M50.9} but at $m_f^{\rm pole}=0.35$.
}
\label{fig:UPP_M50.9_mf}
\end{minipage}
\begin{minipage}{0.51\hsize}
\includegraphics[width=70mm, bb=0 0 345 230]{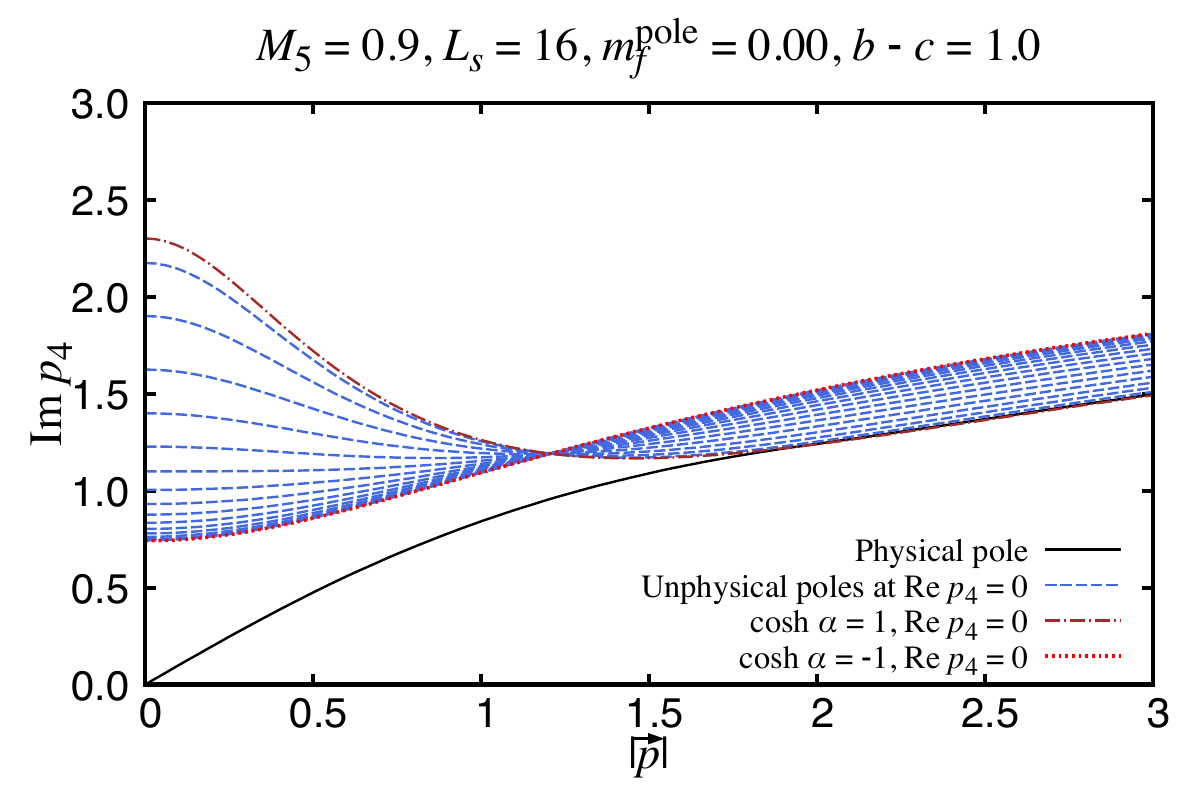}
\caption{
Same as Figure~\ref{fig:UPP_M50.9} but at $L_s=16$.
}
\label{fig:UPP_M50.9_Ls}
\end{minipage}
\end{figure}

Before varying $M_5$ and $b-c$, which play a key role in
determining the region of unphysical pole energies, we briefly show the
results of varying the other parameters $m_f^{\rm pole}$ and $L_s$.
Figure~\ref{fig:UPP_M50.9_mf} shows the result in a massive case
at $m_f^{\rm pole}=0.35$ with the same values of the other parameters as
those in Figure~\ref{fig:UPP_M50.9}.
Compared to Figure~\ref{fig:UPP_M50.9}, only the physical pole is supposed
to depend significantly on the input mass parameter.
The small $m$-dependence of the unphysical poles is compatible with the fact that
the boundaries \eqref{eq:sol_cosha_p1} \eqref{eq:sol_cosha_m1} of the
unphysical poles are independent of $m$.
Therefore, as the physical pole mass $m_f^{\rm pole}$ increases,
it approaches the unphysical pole masses and the dominance of the physical
pole would be lost.

In Figure~\ref{fig:UPP_M50.9_Ls}, we show the dispersion relation at
$L_s=16$.
As described in the previous section, $F_{L_s}$ oscillates in the region satisfying
$-1<\cosh\alpha<1$ with varying $p_4$ and the frequency of the oscillation
is proportional to $L_s$.
The number of unphysical poles has therefore increased to 15.

\begin{figure}[tbp]
\begin{minipage}{0.51\hsize}
\includegraphics[width=70mm, bb=0 0 345 230]{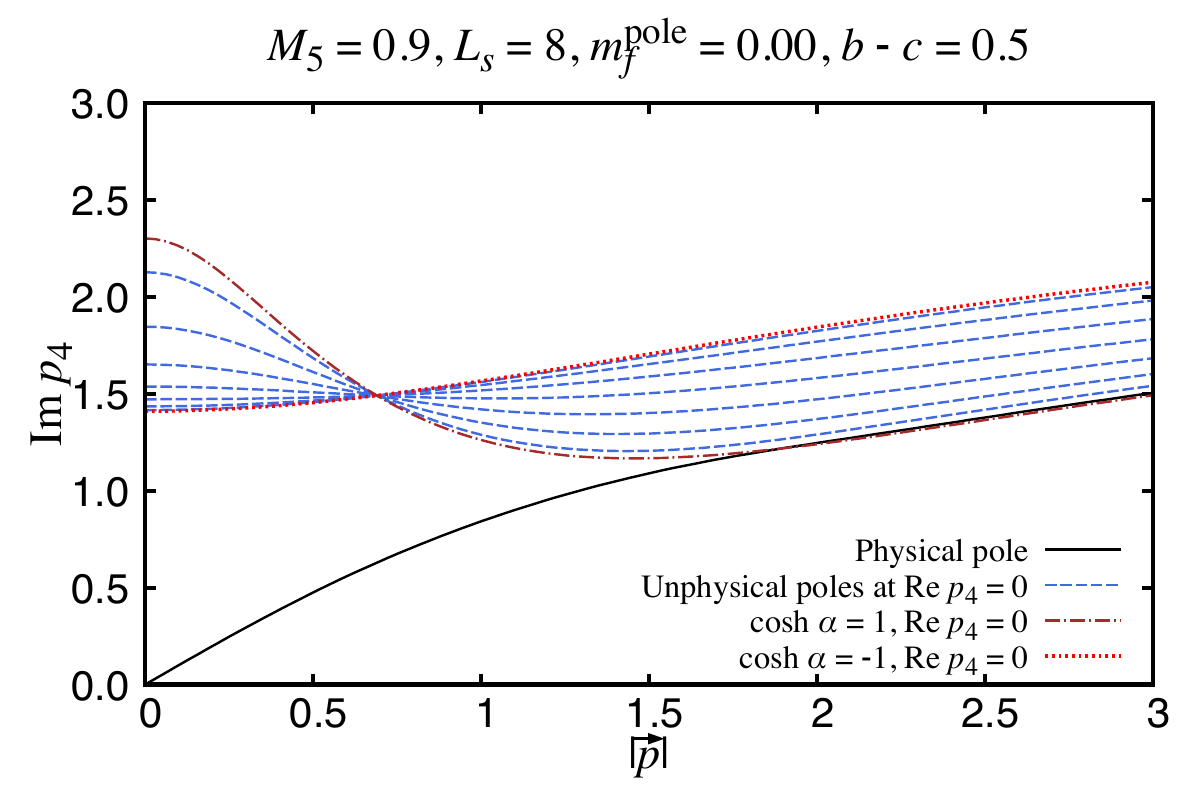}
\caption{
Same as Figure~\ref{fig:UPP_M50.9} but at $b-c=0.5$.
}
\label{fig:UPP_M50.9_c0.5}
\end{minipage}
\begin{minipage}{0.51\hsize}
\includegraphics[width=70mm, bb=0 0 345 230]{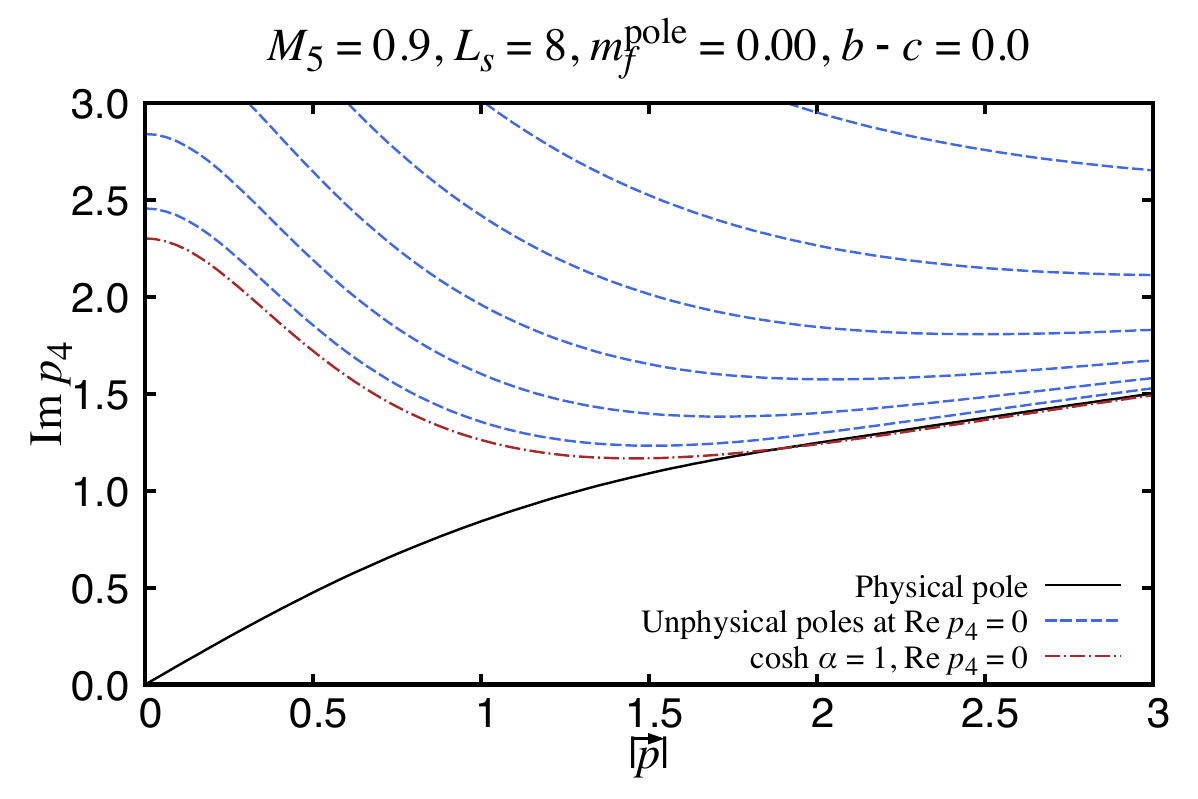}
\caption{
Same as Figure~\ref{fig:UPP_M50.9} but at $b-c=0$.
}
\label{fig:UPP_M50.9_c0.0}
\end{minipage}
\end{figure}

Figure~\ref{fig:UPP_M50.9_c0.5} shows the result at $b-c=0.5$.
In the plot, one bound on the unphysical poles satisfying $\cosh\alpha=-1$
is larger than that for the Shamir type $b-c=1$ and the lightest unphysical
pole mass has been increased to $\sim1.41$,
implying that the contribution of unphysical poles at long distances would
be suppressed more rapidly.
In Figure~\ref{fig:UPP_M50.9_c0.0}, which shows the result at $b-c=0$,
the curve of $\cosh\alpha=-1$ is infinitely large as
\eqref{eq:sol_cosha_m1} indicates.
Thus, small values of $b-c$ make the unphysical modes heavy and would
realize a small contribution of unphysical poles to four-dimensional physics.

\begin{figure}[tbp]
\begin{minipage}{0.51\hsize}
\includegraphics[width=70mm, bb=0 0 345 230]{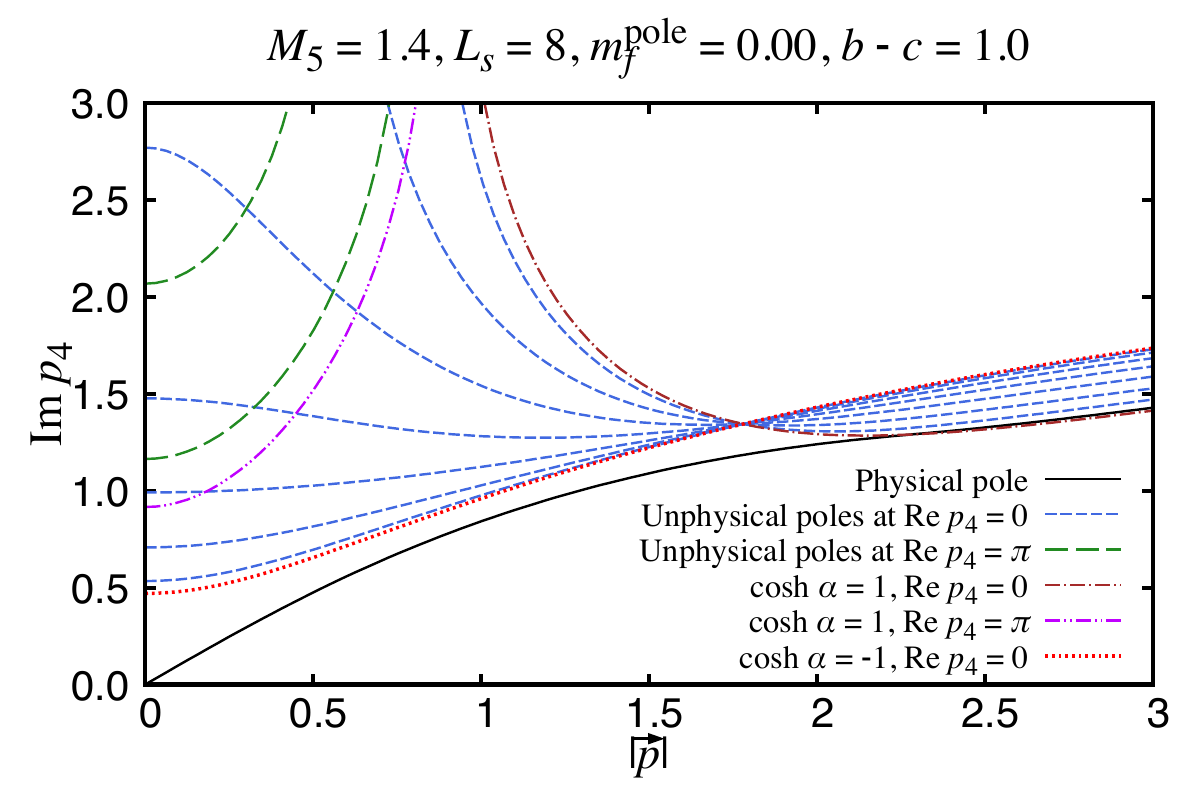}
\caption{
Same as Figure~\ref{fig:UPP_M50.9} but at $M_5 = 1.4$.
}
\label{fig:UPP_M51.4_c1.0}
\end{minipage}
\begin{minipage}{0.51\hsize}
\includegraphics[width=70mm, bb=0 0 345 230]{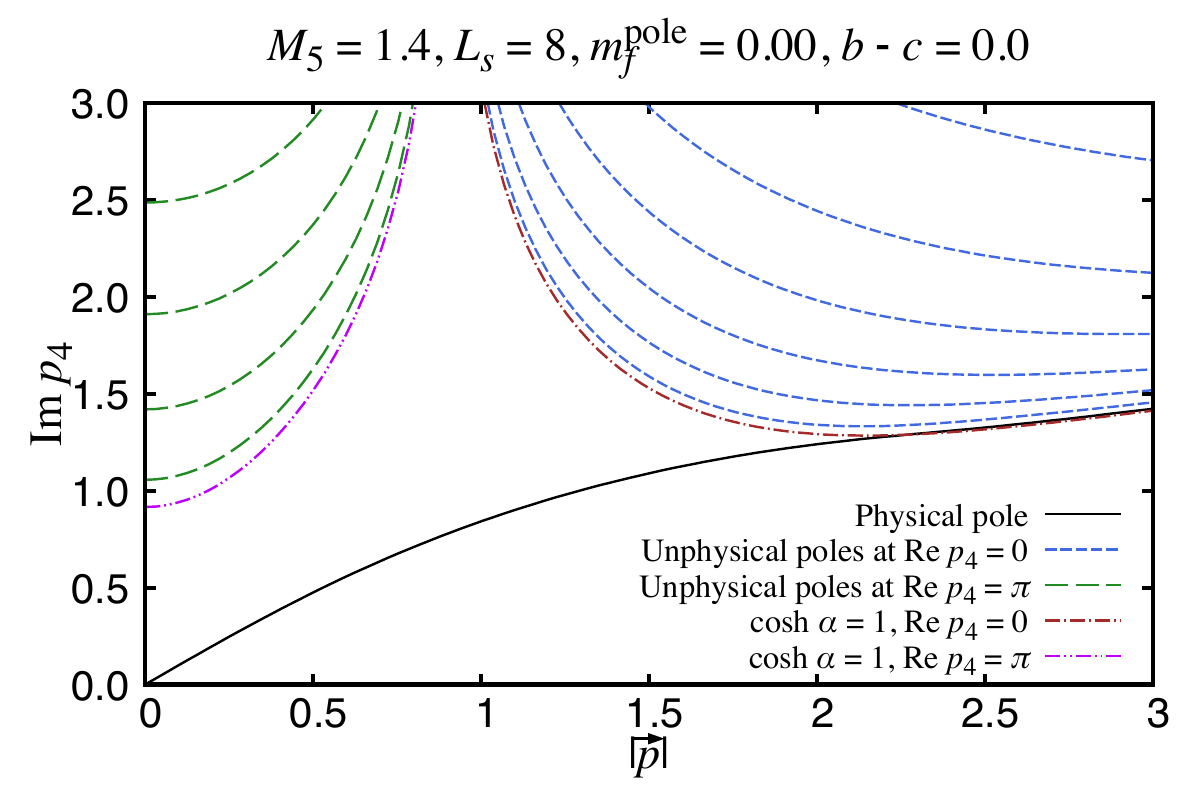}
\caption{
Same as Figure~\ref{fig:UPP_M51.4_c1.0} but at $b-c=0$.
}
\label{fig:UPP_M51.4_c0.0}
\end{minipage}
\end{figure}

So far, we have discussed the case of $M_5=0.9$, in which unphysical poles
are located only on the imaginary axis of $p_4$.
If $M_5>1$, $\alpha$ could be pure imaginary at ${\rm Re}~p_4=\pi$ as
well as at ${\rm Re}~p_4=0$ and therefore some of the unphysical poles
may exist at ${\rm Re}~p_4=\pi$.
Figure~\ref{fig:UPP_M51.4_c1.0} shows the result at $M_5=1.4$.
There are two separated curves of $\cosh\alpha=1$, which blow up at $|\vec p\,|\simeq0.90$.
One of them at smaller spatial momenta (dashed double-dotted curve) is
located at ${\rm Re}~p_4=\pi$.
As suggested in \cite{Liang:2013eoa,Sufian:2016cft}, unphysical poles at
${\rm Re}~p_4=\pi$ may cause unphysical oscillation since the contribution
of a pole at $p_4=p_4^{\rm pole}$ to the quark propagator for the time direction
has a term $\sim\e^{\img p_4^{\rm pole}x_4}$, which is oscillatory unless
${\rm Re}~p_4^{\rm pole}=0$.
In Figure~\ref{fig:UPP_M51.4_c1.0}, the lower bound on the unphysical
pole masses at ${\rm Re}~p_4=0$ ($\cosh\alpha=-1$) is smaller than that at
${\rm Re}~p_4=\pi$ ($\cosh\alpha=1$), indicating that the unphysical contributions
from the former type of poles may be more significant than those from the latter
type of poles.
Figure~\ref{fig:UPP_M51.4_c0.0} shows the result at $b-c=0$.
Since the boundary of $\cosh\alpha=-1$ goes to infinity at $b-c=0$,
there are no unphysical poles on the imaginary axis of $p_4$ at small spatial
momenta ($|\vec p\,|\lesssim0.90$).
All unphysical poles are located at ${\rm Re}~p_4=\pi$ for $|\vec p\,|\lesssim0.90$
and at ${\rm Re}~p_4=0$ for $|\vec p\,|\gtrsim0.90$.

As we have seen, the lower bound on the unphysical pole energies can be increased
by taking $b-c$ smaller when $p_4|_{\cosh\alpha=-1} < p_4|_{\cosh\alpha=1}$.
Since $b-c=1$ satisfies this inequality at zero and small spatial momenta,
this fact implies that we can reduce the contamination of unphysical poles by
decreasing $b-c$ from 1 at least in free field theory.
Non-perturbative effects may slightly change this prospect and this possibility
motivates us to implement a non-perturbative study on the effect of unphysical
poles, which is on-going.

\section{Summary}
\label{sec:summary}

We have shown that the propagator of domain-wall fermions has $L_s-1$ extra
poles.
Since the energy-momentum dispersion relation for unphysical poles is mostly
independent of input physical quark mass, these poles may affect 4D physics
significantly when the input quark mass is comparable to the unphysical pole
masses, which is $O(a^{-1})$.
Examining the dependence on parameters of M\"obius domain-wall fermions,
we demonstrate that $b-c$ should be smaller than 1 for a rapid suppression of
the contamination of unphysical poles at least in free field theory.

It should be also noted that small values of $b-c$ may spoil the approximation
of the sign function because the upper limit on the eigenvalues of the M\"obius
kernel becomes large.
Therefore, we need to tune the parameters taking account of the violation of
the Ginsparg-Wilson relation as well as of the effects of unphysical poles.
A non-perturbative study to tune the parameters of M\"obius domain-wall
fermions is on-going.
The property of unphysical poles is discussed in the full paper \cite{Tomii:2017lyo}
in more detail.

\bibliography{Lattice2017_166_TOMII}

\end{document}